\documentclass[manuscript]{aastex}

\newcommand{\msec}[2]{$#1\mbox{$''\mskip-7.6mu.\,$}#2$}

\newcommand{\myemail}{laurent@mpifr-bonn.mpg.de}

\newcommand{\dechms}[4]{$#1^{\rm h}#2^{\rm m}#3\mbox{$^{\rm s}\mskip-7.6mu.\,$}#4$}

\newcommand{\decdms}[4]{$#1^{\circ}#2'#3\mbox{$''\mskip-7.6mu.\,$}#4$}

\slugcomment{}

\shorttitle{Molecular material associated with $\eta$ Carinae}
\shortauthors{Loinard et al.}

\usepackage{epsfig}
\usepackage{longtable}
\usepackage{amssymb}
\usepackage{lscape}
\usepackage{color}

\begin{document}

\title{Spatial distribution and kinematics of the molecular material associated with $\eta$ Carinae}

\author{Laurent Loinard\altaffilmark{1,2}, 
Tomasz Kami\'nski\altaffilmark{3},\\
Paolo Serra\altaffilmark{4}, 
%Rolf G\"usten\altaffilmark{1},  
Karl M.\ Menten\altaffilmark{1}, 
Luis A.\ Zapata\altaffilmark{2},
and
Luis F.\ Rodr\'{\i}guez\altaffilmark{2}
}

\altaffiltext{1}{Max Planck Institut f\"ur Radioastronomie, Auf dem H\"ugel 69, 53121 Bonn, Germany (\myemail)}

\altaffiltext{2}{Instituto de Radioastronom\'{\i}a y Astrof\'{\i}sica, Universidad
Nacional Aut\'onoma de M\'exico\\ Apartado Postal 3-72, 58090,
Morelia, Michoac\'an, Mexico }

\altaffiltext{3}{ESO, Alonso de C\'ordova 3107, Vitacura, Casilla, 19001, Santiago, Chile}

\altaffiltext{4}{CSIRO Astronomy, Space Science, Australia Telescope National Facility, PO Box 76, Epping, NSW 1710, Australia}

\begin{abstract}
Single-dish sub-millimeter observations have recently revealed the existence of a substantial, chemically peculiar, molecular gas component located in the innermost circumstellar environment of the very massive luminous blue variable star $\eta$ Carinae.  Here, we present 5$''$-resolution interferometric observations of the 1$\rightarrow$0 rotational transition of hydrogen cyanide (HCN) obtained with the Australia Telescope Compact Array (ATCA) toward this star. The emission is concentrated in the central few arcseconds around $\eta$ Carinae and shows a clear 150 km s$^{-1}$ velocity gradient running from west-north-west (blue) to east-south-east (red). Given the extent, location, and kinematics of this molecular material, we associate it with the complex of dusty arcs and knots seen in mid-infrared emission near the center of the Homunculus nebula. Indeed, the shielding provided by this dust could help explain how molecules survive in the presence of the intense UV radiation field produced by $\eta$ Carinae. The dust located in the central few arcseconds around $\eta$ Carinae and the molecular component described here have most likely formed {\it in situ}, out of material expelled by the massive interacting binary system. Thus, $\eta$ Carinae offers us a rare glimpse on the processes leading to the formation of dust and molecules around massive stars that are so relevant to the interpretation of dust and molecule detections at high redshifts.

\end{abstract}


\section{Introduction}

As one of the most massive stellar sources in the Milky Way, $\eta$ Carinae provides us with a unique opportunity to study the intricate evolution of very massive stars and their ejecta. It entered the ``Hall of Fame'' of astronomy in the mid 19th century when it underwent a major outburst, temporarily becoming the second brightest star at visible wavelengths in the entire sky. Known as the Great Eruption, this event led to the ejection of at least ten solar masses of material (and perhaps as much as 40 M$_\odot$; Smith et al. 2003, Gomez et al. 2010) now distributed in a bipolar, vaguely anthropomorphic, nebula called the {\em Homunculus} (Gaviola 1950). From tip to tip, the Homunculus is now about 19$''$ long, with each lobe currently expanding at $\sim$ 700 km s$^{-1}$ (e.g.\ Currie \& Dowling 1999; Davidson 1999; see also Figure 4). A second, less dramatic, brightening of $\eta$ Carinae occurred in the 1890s (Frew et al.\ 2004), and has been interpreted as a normal LBV (S Dor-type) outburst (Humphreys \& Davidson 1999). Ishibashi et al.\ (2003) identified components in the central $\sim$4$''$ of the Homunculus Nebula whose spatio-kinematical characteristics are consistent with an ejection during the 1890s. They nicknamed this smaller structure the {\em little Homunculus} (see their Figure 10), showed that its expansion velocity is about 300 km s$^{-1}$, and concluded that during the 1890s outburst, the stellar mass loss of $\eta$ Carinae increased but in the form of a relatively slow wind. 

The spatial distribution of the dust in $\eta$ Carinae remains somewhat unclear. Optical images obtained with the Hubble Space Telescope certainly show that dust is present thoughout the nebula as well as in the {\em equatorial skirt} region (e.g.\ Davidson \& Humphreys 1997), but they provide little quantitative information on the overall density distribution of the dust. Mid-infrared observations (e.g.\ Smith et al.\ 2002; Chesneau et al.\ 2005) revealed bright emission, interpreted as thermal dust radiation, emanating from the central few arcseconds around $\eta$ Carinae. At high resolution (see Figure 1 in Smith et al.\ 2002), the emission is resolved into an intricate complex of knots, filaments, and arcs roughly oriented along the equatorial plane of the Homunculus. The overall aspect of this mid-infrared emission is reminiscent of the shape of a butterfly, and as been referred to as the {\it Butterfly Nebula} (Chesneau et al.\ 2005). Spatially, this dust component occupies the same region as the little Homunculus, suggesting that a relation might exist between the two. As discussed by Chesneau et al.\ (2005), however, this relation is presumably complex, and difficult to assess in the absence of kinematic information on the dust component.

Several molecules are also known to be associated with the Homunculus. Molecular hydrogen as traced by its 2.12 $\mu$m emission (Smith 2006) as well as CH and OH detected through their ultraviolet absorption lines (Verner et al.\ 2005) are known to trace the thin outer layer of the Homunculus. Other simple molecules\footnote{Smith et al.\ (2006) reported on the detection of the ammonia (3,3) inversion line in $\eta$ Carinae, but more recent observations --to be published in a forthcoming paper-- show that the detected line is not NH$_3$(3,3) at  23.870 GHz, but rather the recombination line H81$\beta$ at 23.861 GHz.} (CO, CN, HCO$^+$, HCN, HNC, and N$_2$H$^{+}$, as well as several of their $^{13}$C isotopologues) were detected through their sub-millimeter rotational emission lines by Loinard et al.\ (2012). These observations collected with the 12 meter single-dish Atacama Pathfinder EXperiment telescope (APEX; G\"usten et al.\ 2006) have an angular resolution of order 15$''$, precluding any detailed mapping. The relative intensities of the detected CO transitions, however, suggest that the emission is significantly more compact than the APEX beam, and concentrated near the very center of the Homunculus, within a few arcseconds of $\eta$ Carinae itself. Furthermore, while the H$_2$, CH, and OH lines trace the $\sim$ 700 km s$^{-1}$ expansion of the Homunculus, the integrated line profiles of the sub-millimeter molecular transitions are only about 200 km s$^{-1}$ wide. This clearly points to a different spatial distribution for H$_2$, CH, and OH on the one hand, and the molecules detected via their sub-millimeter transitions, on the other. 

\section{Observations and Results}

Here, we report on interferometric observations of the HCN(1-0) rotational line at 88.6316 GHz obtained with the Australia Telescope Compact Array (ATCA) in its H75 configuration on 2014, 07--08 July. This configuration provides baselines with lengths between 30 and 90 m. The Compact Array Broadband Backend (CABB) was used to record a total bandwidth of 2 GHz, with a spectral resolution of 1 MHz --corresponding to 3.4 km s$^{-1}$ at this frequency. The frequency response (bandpass) of the system was determined from a 20-minute integration on the bright quasar 1253--055, while the absolute flux scale was defined using an observation of Uranus obtained at the beginning of the observations. We estimate the absolute flux accuracy to be 15\%. The phase and amplitude gains were monitored using observations of the quasar 1045--62 interspersed every 15 minutes with those of $\eta$ Carinae. 

The data calibration was performed in Miriad (Sault et al.\ 1995) using the standard protocol for high frequency observations as described in Stevens (2012).\footnote{\scriptsize http://www.narrabri.atnf.csiro.au/people/Jamie.Stevens/CX208/tutorial/basic$\_$continuum$\_$tutorial.pdf} From the calibrated visibilities, a continuum image was constructed using a weighting scheme of the visibilities intermediate between natural and uniform (Robust = 0 in Miriad). It yields an angular resolution of \msec{5}{9} $\times$ \msec{4}{4}, at P.A.\ = --87$^\circ$. This initial image revealed the presence of a bright ($\sim$ 20 Jy) compact source associated with $\eta$ Carinae, and was used as a model to self-calibrate the visibilities in phase and improve upon the standard calibration described above. A new continuum image was produced after applying the self-calibration gains. In this final continuum image, the 88.6 GHz continuum of $\eta$ Carinae was measured to be 20.6 $\pm$ 3.1 Jy. Abraham et al.\ (2014) recently reported multi-frequency continuum observations of $\eta$ Carinae obtained with ALMA in November 2012. They measure a flux of 28.4 Jy at 92.5 GHz, presumably with an uncertainty of order 10\%. Using the ALMA fluxes at the other frequencies, we interpolate an ALMA flux of 26.6 $\pm$ 3.0 Jy at 88.6 GHz, which agrees within 1$\sigma$ with our own determination. We note, in addition, that a large fraction of the continuum flux of $\eta$ Carinae at $\sim$90 GHz is known to be of free-free origin (e.g.\ Abraham et al.\ 2014) and to be variable, with the flux being minimum near periastron and maximum near apastron (White et al.\ 2005). Our own observations were obtained at epoch 2014.52, very near periastron (2014.59; Damineli et al.\ 2008; Teodoro et al.\ 2016). The ALMA data reported by Abraham et al.\ (2014), on the other hand, correspond to epoch 2012.84, about 1.67 years before periastron. It is, therefore, entirely natural that the flux in the ATCA observations reported here should be 25\% smaller than that measured by Abraham et al.\ (2014) one and a half year earlier.

In addition to the continuum image, we produced a continuum-free data cube centered on the HCN(1-0) line. The continuum subtraction was performed directly on the visibilities, interpolating the spectrum around the HCN(1-0) line with a first order polynomial. Figure 1 shows the correlated intensity (in Jy) present in the entire field mapped by ATCA. There is an obvious similarity between the line profile in Figure 1 and the molecular lines detected with APEX (see Figure 1 in Loinard et al.\ 2012). In particular, the four distinct velocity components identified in the APEX spectra at $v_{LSR}$ $\sim$ $-$76, $-$9, $+$30, and $+$63 km s$^{-1}$ are also clearly seen in the spatially integrated ATCA spectrum, demonstrating that the two instruments trace the same gas. It is instructive to compare the relative intensities of the APEX and ATCA spectra. The peak flux of the HCN(1-0) spectrum in the ATCA observations is about 0.2 Jy (Figure 1). A different HCN line (corresponding to the 4$\rightarrow$3 transition) was observed with APEX, but under the reasonable assumption that the lines are optically thick and trace warm gas, the line brightness is expected to be equal to the kinetic temperature of the gas independently of the transition considered. This very behavior is indeed observed for the CO lines in the APEX observations (Loinard et al.\ 2012). The main beam brightness temperature of the HCN(4-3) line detected with APEX was T$_b$ $\sim$ 0.1 K (Loinard et al.\ 2012) in a $\theta_b$ = \msec{17}{5} beam. This corresponds to a flux density $S$ = ${2 k \nu^2 T_b \Omega_b / c^2}$ $\sim$ 0.2 Jy, in excellent agreement with the flux recovered in our ATCA data (in the previous formula, $k$, $c$, and $\nu$ have their usual meaning, while $\Omega_b$ = $\pi \theta_b^2/4 \ln(2)$ is the solid angle corresponding to the telescope beam). This shows that the entire single dish flux is recovered in the interferometer data and confirms that the molecular emission is significantly more compact than the single dish beam. 

Channel maps corresponding to the velocity range over which HCN(1-0) emission is detected are shown in Figure 2. The emission is found to be restricted to the central few arcseconds around $\eta$ Carinae itself, although it appears to be slightly elongated in several velocity channels. In addition, the location of the emission varies as a function of velocity. The most negative velocity component (at $v_{LSR}$ $\sim$ $-$76 km s$^{-1}$) corresponds to a feature located about 2$''$ almost exactly due west from $\eta$ Carinae. The emission at $v_{LSR}$ $\sim$ $-$10 km s$^{-1}$ is located very near $\eta$ Carinae itself, whereas the emission at positive $LSR$ velocities is mostly found to the south and east of the stellar system. There is an apparent absorption feature at $v_{LSR}$ $\sim$ --60 km s$^{-1}$ (Figures 1 and 2). With the information currently available, it is impossible to tell whether this absorption is real or an artifact of the imperfect $(u,v)$ coverage of our interferometric observations. 

The integrated intensity HCN(1-0) map (i.e. the zeroth moment of the emission) is shown as contours in Figure 2 and 3. It is slightly elongated in the east-west direction, with a deconvolved size of order a few arcseconds (it is difficult to provide a more accurate estimate, given the synthesized beam of the present observations). In the left panel of Figure 3, we show the velocity field (first moment) traced by the HCN(1-0) emission. In agreement with the previous paragraph, we find a velocity gradient running from west-north-west (where $v_{LSR}$ $\sim$ $-$90 km s$^{-1}$) to east-south-east (where $v_{LSR}$ $\sim$ $+$70 km s$^{-1}$). At the position of $\eta$ Carinae itself, the weighted velocity is $v_{LSR}$ $\sim$ $+$22 km s$^{-1}$. This is significantly red-shifted from the nominal systemic velocity of $\eta$ Carinae ($v_{LSR}$ $\sim$ $-$20 km s$^{-1}$; Smith 2004). In fact, adopting the latter value for the systemic velocity of $\eta$ Carinae leads to the conclusion that it contains very little blue-shifted molecular gas: only the outer western layer (traced by the emission in the component at $v_{LSR}$ $\sim$ $-$76 km s$^{-1}$) appears to be expanding toward us. The right panel of Figure 3 shows the second moment of HCN(1-0) emission, revealing the velocity dispersion as a function of position. Across most of the source, the dispersion is of order 40 km s$^{-1}$ because this is the typical linewidth of each of the three individual velocity components identified in Figure 1 at $-$9, $+$30, and $+$63 km s$^{-1}$ (Loinard et al.\ 2012). Only to the east of $\eta$ Carinae is the dispersion larger (reaching 60 km s$^{-1}$). This region corresponds to the location of the narrow velocity component $v_{LSR}$ $\sim$ $-$76 km s$^{-1}$ but also coincides with emission at positive velocities (see Figure 2). 

We should emphasize that the angular resolution of our HCN(1-0) observations is only about 5$''$ (i.e.\ the observed images correspond to the convolution of the true emission with a 5$''$ beam). The fairly smooth aspect of the images shown in Figures 2 and 3 are, largely, due to this modest angular resolution and should not be interpreted as evidence that the molecular gas in $\eta$ Carinae is smoothly distributed. Indeed, the HCN(1-0) spectrum (Figure 1) suggests the existence of several distinct velocity components. At high angular resolution, the distribution of the molecular gas in $\eta$ Carinae is likely to be highly structured. For the same reason, the true extent of the molecular gas distribution is, undoubtedly, much smaller than the size of the lower contours in the moment zero image. In Figure 4, we overplot our HCN(1-0) moment zero image on a false-color Hubble Space Telescope (HST) image of $\eta$ Carinae (see the caption of Figure 4 for details). This comparison demonstrates that the molecular gas detected via the HCN(1-0) line is associated only with the central region of $\eta$ Carinae, and not with the entire Homunculus. Accounting for the foregoing warnings regarding the modest angular resolution of the HCN observations, Figure 4 should emphatically not be interpreted as evidence that the molecular gas occupies the entire region enclosed in the lowest contour level. It is instead concentrated in the central few arcseconds around $\eta$ Carinae itself.

\section{Discussion and Perspectives}

The ATCA observations presented above confirm the suspicion based on the single dish APEX observations that the millimeter and sub-millimeter molecular lines trace material associated with the central few arcseconds around $\eta$ Carinae. The HCN(1-0) line profile covers the LSR velocity range between $-$100 and $+$100 km s$^{-1}$ (Figure 1) that lies within the velocity range associated with the little Homunculus by Ishibashi et al.\ (2003). Thus, both the spatial and the kinematical properties of the molecular emission appear, at first sight, to be similar to those of the little Homunculus. Such an association, however, can be discarded on the basis of the first moment map (Figure 3) because the velocity gradient seen in the molecular emission (red to the south, blue to the north) appears to be in the opposite direction of that in the little Homunculus (red to the north and blue to the south; e.g.\ Ishibashi et al.\ 2003).

Instead, we associate the molecular component reported here with the warm dust detected in the central few arcseconds of the Homunculus through mid-infrared observations (Smith et al.\ 2002; Chesneau et al.\ 2005). This dusty structure located in the central $\sim$ 4$''$ $\times$ 2$''$ of the Homunculus exhibits a kinematics (negative velocities up to about $-$100 km s$^{-1}$ to the North; positive velocities up to about $+$100 km s$^{-1}$ to the South; Smith 2006) as the molecular emission. Certainly, since the formation mechanisms of dust and molecules are intimately intertwined (e.g.\ Tielens 2010), an association between the molecular component in $\eta$ Carinae and a warm dust structure is far from unexpected. This association would, indeed, help explain how molecules (such as CO, HCN, HNC, etc.) with dissociative energies $\lesssim$ 10 eV can survive in the presence of the intense UV radiation field provided by the central massive stars in $\eta$ Carinae: in this scenario, the dust grains provide shielding to the molecules from the destructive UV photons. The presence of dust in the Homunculus nebula around $\eta$ Carinae has long been interpreted as evidence for the occurence of a rapid {\it in situ} dust formation mechanism (e.g.\ Westphal \& Neugebauer 1969; Hackwell et al.\ 1986; Smith et al.\ 2002). Our detection of a molecular gas component associated with the central few arcseconds around $\eta$ Carinae demonstrates that molecule formation similarly occurred rapidly and {\it in situ}. 

There is some debate about the exact nature and the origin of the dust seen in the central region of the Homunculus. Morris et al.\ (1999) associate it with a torus pre-dating the Great Eruption, that could have been responsible for ``pinching'' the waist of the Homunculus. Hony et al.\ (2001), on the other hand, argued that it might trace overlapping rings similar to those found in SN 1987a (Burrows et al.\ 1995). The most detailed observations (Smith et al.\ 2002; Chesneau et al.\ 2005) suggest that this dust traces a warm ($\sim$ 300 K) torus highly disrupted by the winds and mass ejections from $\eta$ Carinae. Whether or not this torus pre-dates the Great Eruption remains unclear. Artigau et al.\ (2011) measured proper motions in the butterfly nebula using multi-epoch near-infrared observations. They found evidence for expansion that could be traced back to {\em both} the Great Eruption of the 1840s and the less spectacular outburst of the 1890s. High-resolution, sub-arcsecond molecular observations will be key to better constraining the kinematics of the central dusty structure in $\eta$ Carinae, and would help constrain its nature and origin. ALMA will, evidently, be the instrument of choice to collect such high-resolution observations.

The spectral observations of the light echoes of the Great Eruption (Prieto et al.\ 2014) indicate a very low temperature, $<$ 4500 K, for the pseudo-photosphere some 300 days after the peak brightness. At these temperatures, molecules could easily have formed and indeed the echo spectra include strong electronic bands of CN. If we assume that the molecular gas traced by HCN was created in the late phases of the Great Eruption, it would have to reside in a sort of a dusty torus that is stable against the outflows sweeping out the material in the polar directions. The putative disrupted torus traced by mid-infrared observations would be a natural candidate for this role, and the association --revealed by our observations-- of the molecular material with the central dusty structure might prove highly relevant. We note that similar dusty tori were postulated to exist around post-AGB stars to explain mixed-chemistry objects (Cohen et al.\ 1999). 

In the future, it will also be important to complete the inventory of molecular species in the Homunculus to relate it with the chemical properties of the dust. Chesneau et al.\ (2005) modelled the 7--14 $\mu$m spectrum of the central few arcseconds around $\eta$ Carinae (see their Figure 11) in terms of dust grain composition. They confirm the presence of non-silicate dust around $\eta$ Carinae, as had previously been proposed by Mitchell \& Robinson (1978) and Morris et al.\ (1999), largely to explain the absence of a silicate feature at 18 $\mu$m in the spectrum of $\eta$ Carinae. Chesneau et al.\ (2005) showed further that the 7--14 $\mu$m spectrum can be explained as a mixture of amorphous olivine (MgFeSiO$_4$) and corundum (Al$_2$O$_3$). The peak of this putative corundum component is located about 1$''$ south-east of $\eta$ Carinae itself, where the positive velocity components seen in our HCN(1-0) spectra is located.\footnote{In contrast, the dust associated with the Weigelt blobs, located north of $\eta$ Carinae, appears to be dominated by silicates (Chesneau et al.\ 2005).} It would be very interesting to search for simple aluminum-bearing molecules (AlO, AlH, etc.) in the gas phase as they are likely part of the chemical networks leading to the formation of corundum (e.g.\ Kaminski et al.\ 2016). In parallel, it would be interesting to also search for molecules in stars similar to $\eta$ Car in our Galaxy (e.g.\ P Cygni or V4650 Sagittarii). 

\section{Conclusions}

In this paper, we reported on $\sim$ 5$''$ interferometric mapping of the 3-mm continuum and HCN(1-0) molecular line emission toward $\eta$ Carinae. The molecular emission is concentrated in the central few arcseconds around $\eta$ Carinae itself, and exhibits a clear 150 km s$^{-1}$ velocity gradient across the source. We argue that this molecular material is associated with the warm dust detected in these same regions via its mid-infrared emission, and that both the molecules and the dust have formed {\it in situ} out of material ejected by the stars over the {\scriptsize $\lesssim$} 180 years that elapsed since the Great Eruption of the 1840s. Thus, $\eta$ Carinae provides us with a rare instance of a very massive star around which the production of dust and molecules can be studied in detail as it is on-going. This is potentially highly relevant for the interpretation of the detection of large amounts of dust and molecular gas at high redshift, since only very massive stars could contribute to the chemical enrichment of the Universe at the early cosmological epochs probed by these measurements  (Carilli \& Walter 2013; Dunne et al.\ 2004). To make further progress, sub-arcsecond molecular observations of $\eta$ Carinae with ALMA will be required. 

\acknowledgments
L.L., L.A.Z., and L.F.R.\  acknowledge the financial support of DGAPA, UNAM, and CONACyT, M\'exico. L.L.\  is indebted to the Alexander von Humboldt Stiftung for financial support. We acknowledge useful discussions with Jammie Stevens, James Urquhart, and Ted Gull. The Australia Telescope Compact Array is part of the Australia Telescope National Facility which is funded by the Australian Government for operation as a National Facility managed by CSIRO.)

\begin{figure*}[!ht]
\begin{center}
\includegraphics[width=1.\linewidth,trim=1.0cm 0.0cm 0.0cm 0.0cm]{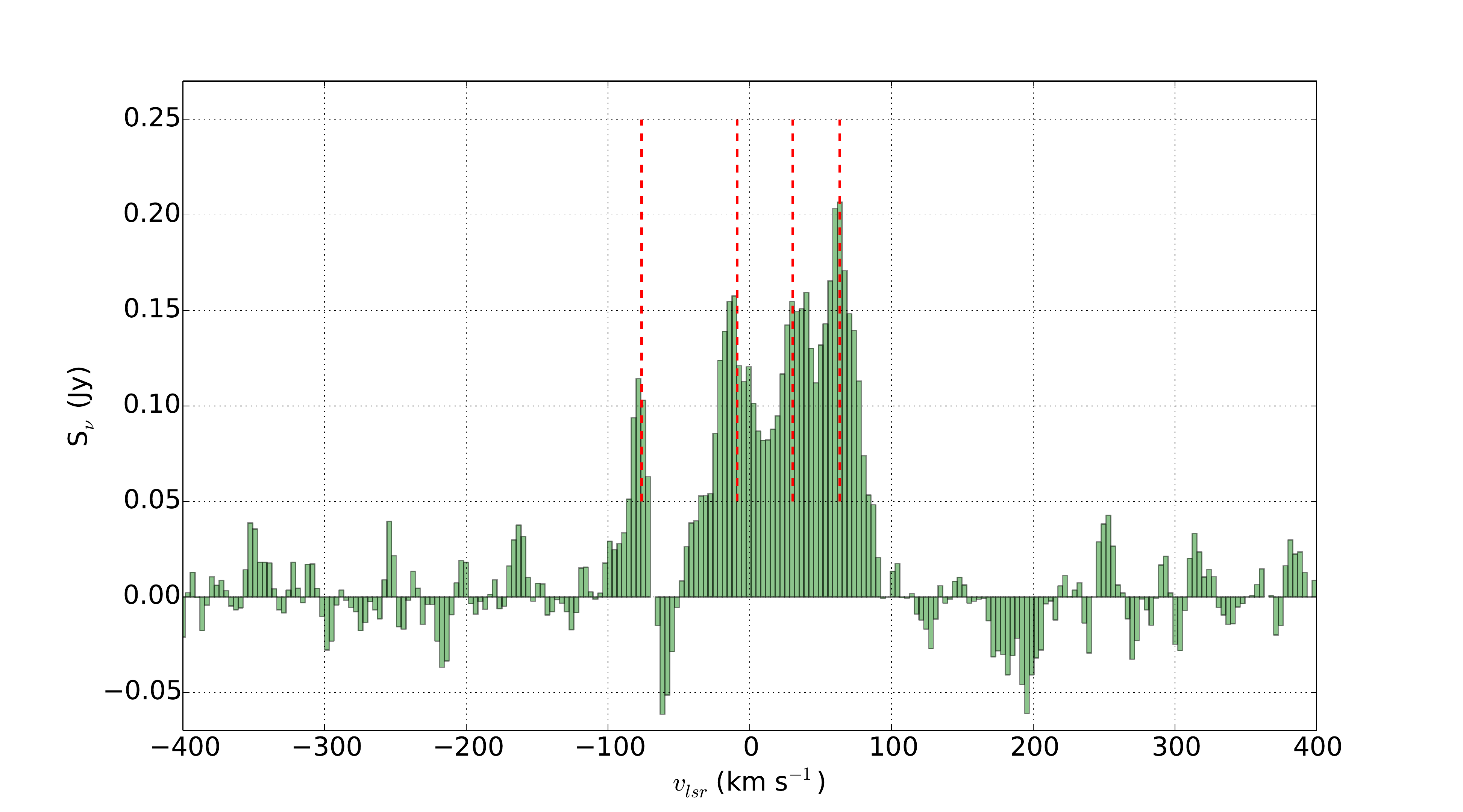}
\end{center}
\caption{Integrated HCN(1-0) spectrum in the ATCA field. The vertical red dashed lines show the velocity components identified in the single dish HCN(4-3) spectrum published by Loinard et al.\ (2012). Here, and in the rest of the paper, the velocities are expressed in the {\it LSR} reference frame. For the direction of $\eta$ Carinae, $v_{hel}$ = $v_{LSR}$ + 11.6 km s$^{-1}$.}
\label{fig:spec}
\end{figure*}

\begin{figure*}[!ht]
\begin{center}
\includegraphics[width=0.9\linewidth,angle=-90,trim=4.0cm 4.0cm 0.0cm 0.0cm]{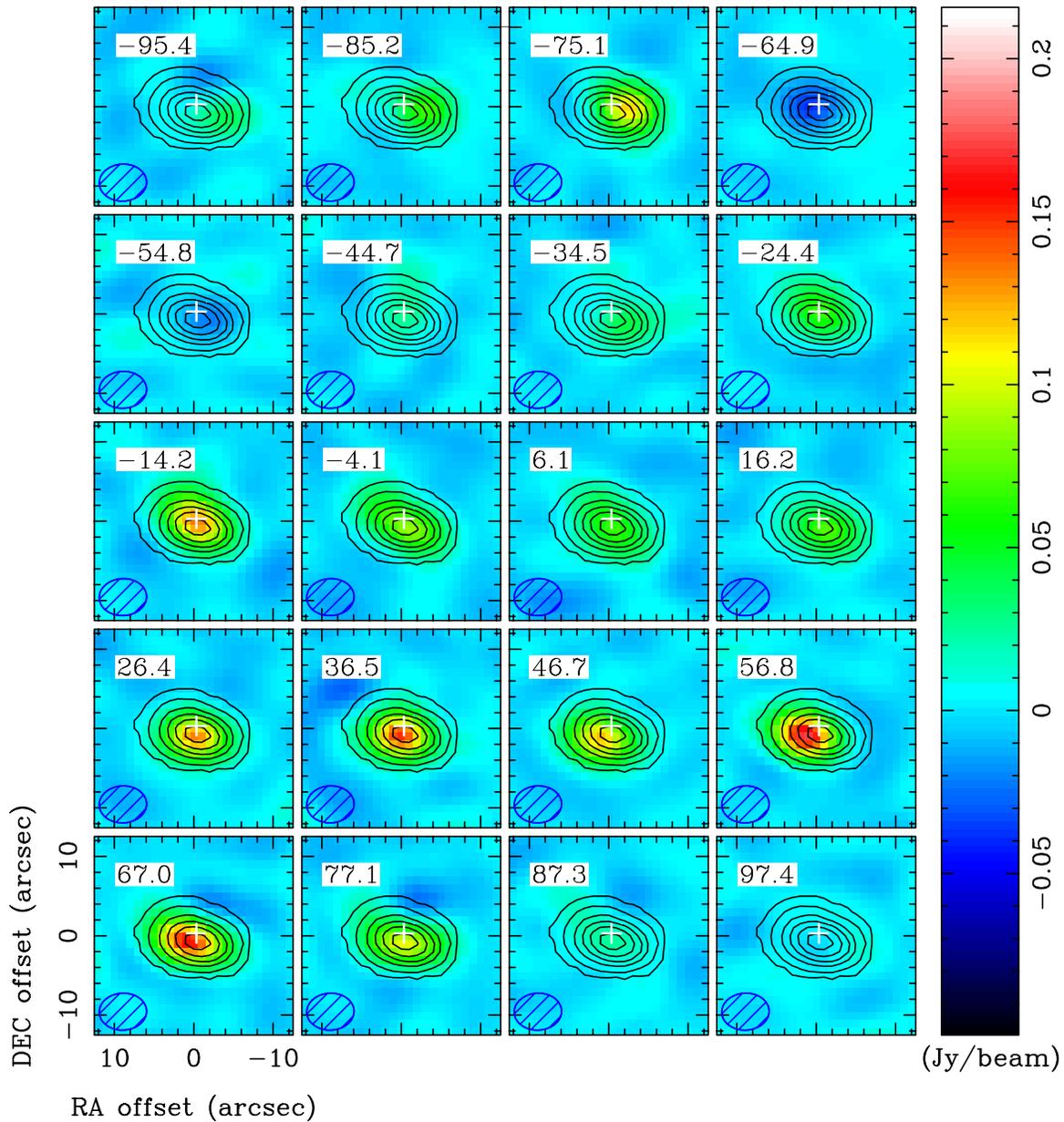}
\end{center}
\caption{HCN(1-0) channel maps. The {\it LSR} velocity is indicated in each panel, and the beam is shown at the bottom left. The coordinates are relative to $\alpha_{ICRF}$ = \dechms{10}{45}{03}{591}, $\delta_{ICRF}$ = \decdms{-59}{41}{04}{26}, and the location of $\eta$ Carinae itself is shown by the white + sign. The noise level in each channel map is about 3.7 mJy beam$^{-1}$. The black contours correspond to the integrated intensity HCN(1-0) map, and are drawn at 1, 10, 30, 50, 70, and 90\% of the peak. }
\label{fig:channels}
\end{figure*}

\begin{figure*}[!ht]
\begin{center}
\includegraphics[width=0.4\linewidth,angle=-90,trim=2.0cm 4.0cm 0.0cm 0.0cm]{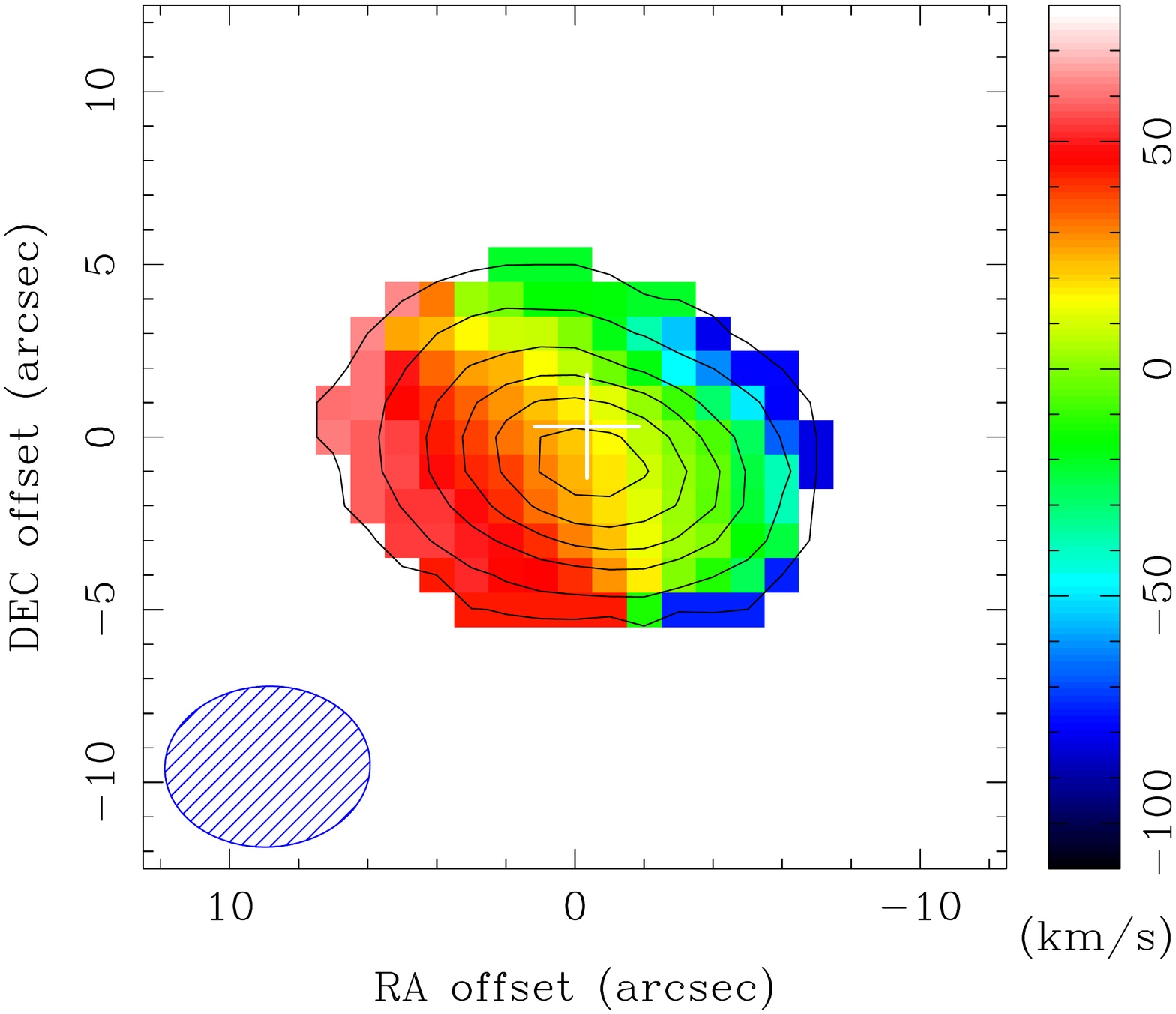}
\includegraphics[width=0.4\linewidth,angle=-90,trim=2.0cm 4.0cm 0.0cm 0.0cm]{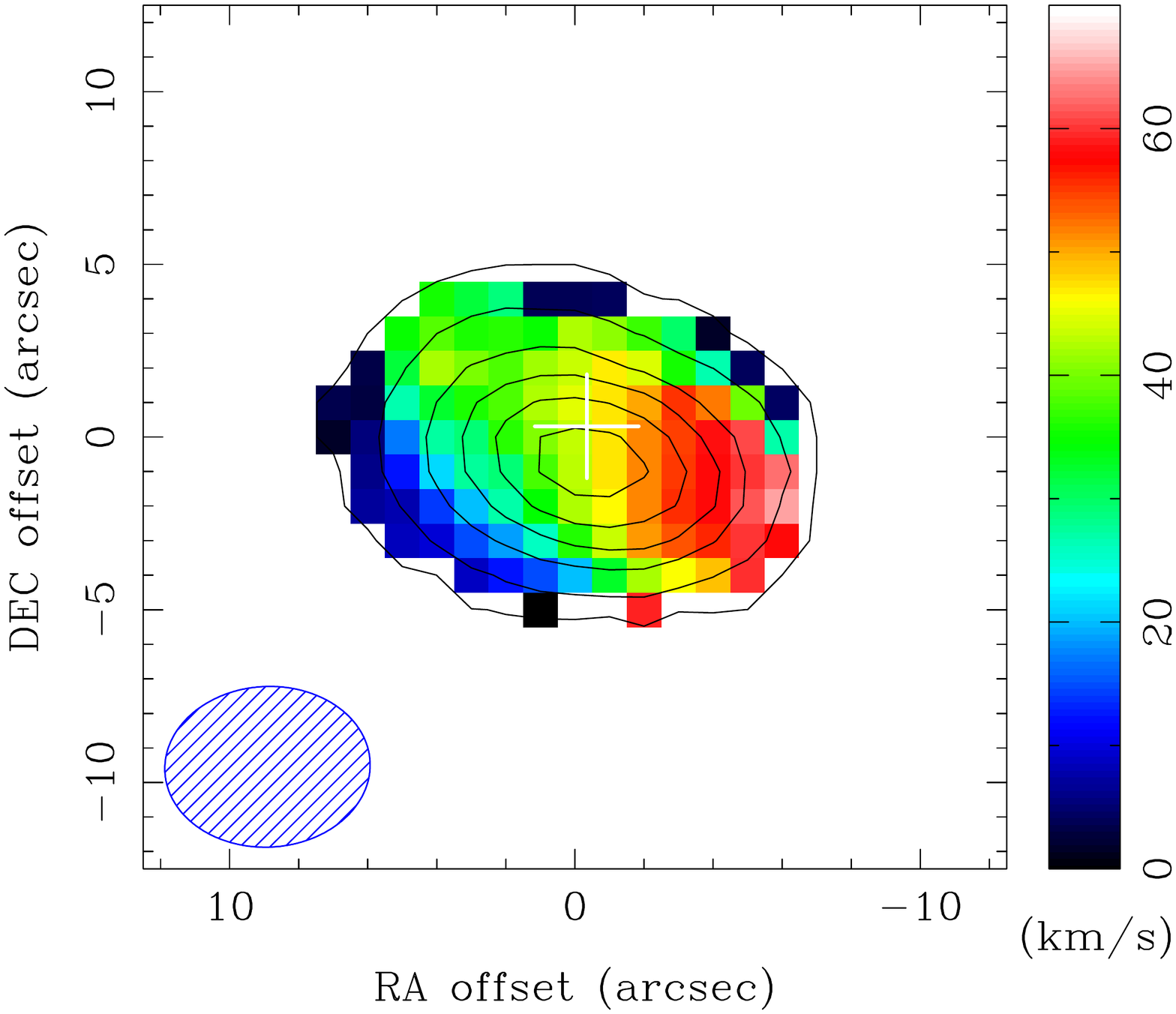}
\end{center}
\caption{HCN(1-0) moment maps in $\eta$ Carinae. The contours in both panels indicate the moment zero map of HCN(1-0) and are the same as in Figure 2. (left:) The colors show the first moment map, corresponding to the velocity field, as indicated by the color wedge to the right of the panel. (right:) The colors show the second moment map, corresponding to the line width, as indicated by the color wedge to the right of the panel.}
\label{fig:moments}
\end{figure*}

\begin{figure*}[!ht]
\begin{center}
\includegraphics[width=0.9\linewidth,trim=0.0cm 0.0cm 0.0cm 0.0cm]{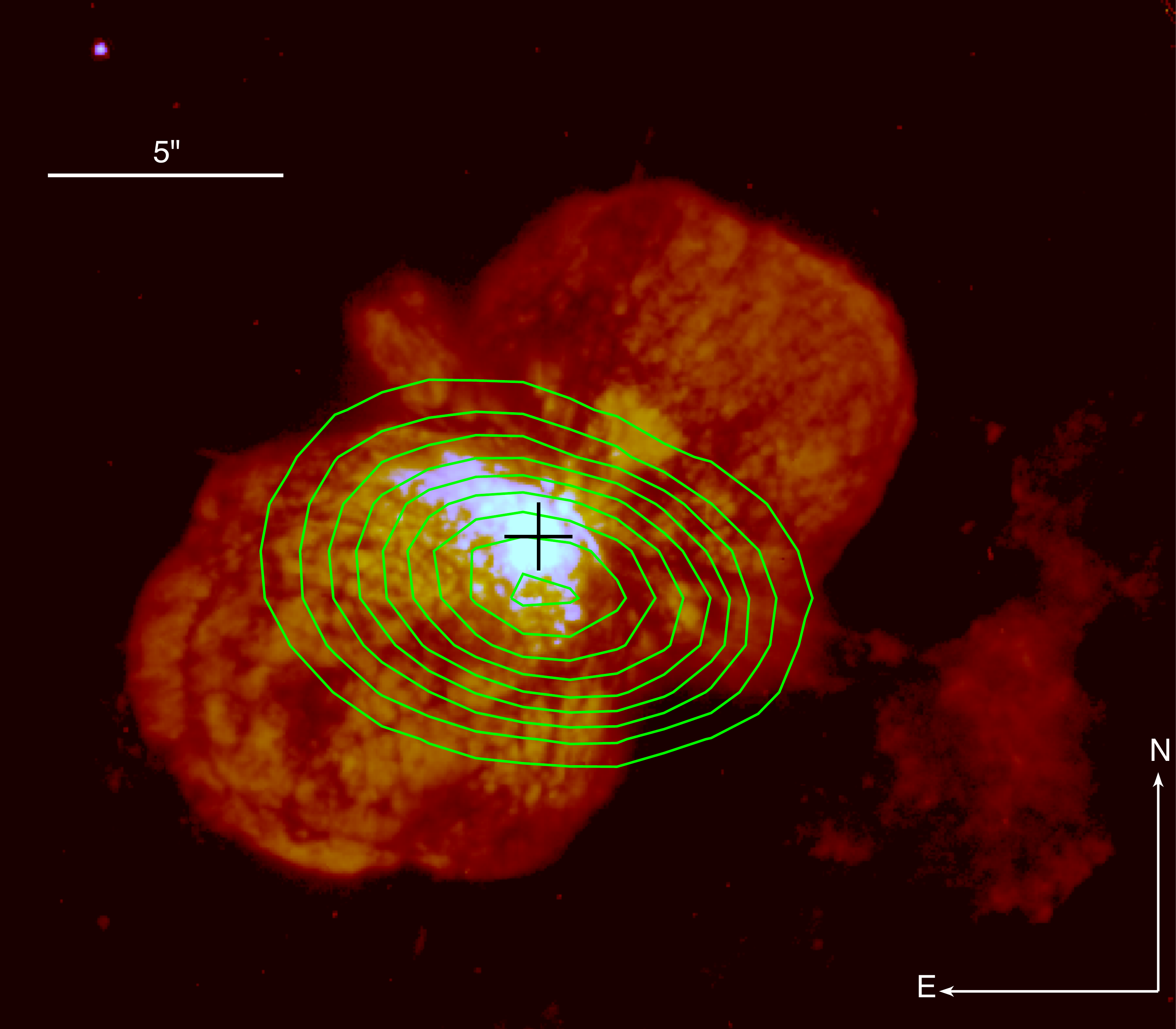}
\end{center}
\caption{HCN(1-0) moment zero map overlaid over a false color HST image. The contours are the same as in Figure 1 but contain one extra contour at 98\% of the peak. The color rendition is a combination of WFPC2 images in the F658N (red) and F631N (blue/white) filters obtained on 2003-02-12. The absolute astrometric calibration of the HST observations is only accurate to about 1$''$. Roughly ten years have elapsed between the HST and ATCA observations shown in this figure; given its expansion speed, the optical Homunculus at the date of the ATCA observations must have been nearly 1$''$ longer (along the major axis direction) that the 18$''$ it measures on the HST image shown here.}
\label{fig:moments}
\end{figure*}


\begin{thebibliography}{}

\bibitem[Abraham et al.(2014)]{2014ApJ...791...95A} Abraham, Z., Falceta-Gon{\c c}alves, D., \& Beaklini, P.~P.~B.\ 2014, \apj, 791, 95 

\bibitem[Artigau et al.(2011)]{2011AJ....141..202A} Artigau, {\'E}., Martin, J.~C., Humphreys, R.~M., et al.\ 2011, \aj, 141, 202 

\bibitem[Burrows et al.(1995)]{1995ApJ...452..680B} Burrows, C.~J., Krist, J., Hester, J.~J., et al.\ 1995, \apj, 452, 680 

\bibitem[Carilli \& Walter(2013)]{2013ARA&A..51..105C} Carilli, C.~L., \& Walter, F.\ 2013, \araa, 51, 105 

\bibitem[Chesneau et al.(2005)]{2005A&A...435.1043C} Chesneau, O., Min, M., Herbst, T., et al.\ 2005, \aap, 435, 1043 

\bibitem[Cohen et al.(1999)]{Cohen1999} Cohen, M., Barlow, M.~J., Sylvester, R.~J., et al.\ 1999, \apjl, 513, L135 

\bibitem[Currie \& Dowling(1999)]{1999ASPC..179...72C} Currie, D.~G., \& Dowling, D.~M.\ 1999, Eta Carinae at The Millennium, 179, 72 

\bibitem[Damineli et al.(2008)]{2008MNRAS.384.1649D} Damineli, A., Hillier, D.~J., Corcoran, M.~F., et al.\ 2008, \mnras, 384, 1649 

\bibitem[Davidson(1999)]{1999ASPC..179....6D} Davidson, K.\ 1999, Eta Carinae at The Millennium, 179, 6 

\bibitem[Davidson \& Humphreys(1997)]{1997ARA&A..35....1D} Davidson, K., \& Humphreys, R.~M.\ 1997, \araa, 35, 1 

\bibitem[Dunne et al.(2004)]{2004NewAR..48..611D} Dunne, L., Morgan, H., Eales, S., Ivison, R., \& Edmunds, M.\ 2004, \nar, 48, 611 

\bibitem[Frew(2004)]{2004JAD....10....6F} Frew, D.~J.\ 2004, Journal of Astronomical Data, 10,  

\bibitem[Gaviola(1950)]{1950ApJ...111..408G} Gaviola, E.\ 1950, \apj, 111, 408 

\bibitem[Gomez et al.(2010)]{2010MNRAS.401L..48G} Gomez, H.~L., Vlahakis, C., Stretch, C.~M., et al.\ 2010, \mnras, 401, L48 

\bibitem[G{\"u}sten et al.(2006)]{2006A&A...454L..13G} G{\"u}sten, R., Nyman, L.~{\AA}., Schilke, P., et al.\ 2006, \aap, 454, L13 

\bibitem[Hackwell et al.(1986)]{1986ApJ...311..380H} Hackwell, J.~A., Gehrz, R.~D., \& Grasdalen, G.~L.\ 1986, \apj, 311, 380 

\bibitem[Hony et al.(2001)]{2001A&A...377L...1H} Hony, S., Dominik, C., Waters, L.~B.~F.~M., et al.\ 2001, \aap, 377, L1 

\bibitem[Humphreys \& Davidson(1999)]{1999ASPC..179..216H} Humphreys, R.~M., \& Davidson, K.\ 1999, Eta Carinae at The Millennium, 179, 216 

\bibitem[Ishibashi et al.(2003)]{2003AJ....125.3222I} Ishibashi, K., Gull, T.~R., Davidson, K., et al.\ 2003, \aj, 125, 3222 

\bibitem[Kami{\'n}ski et al.(2016)]{2016arXiv160405641K} Kami{\'n}ski, T., Wong, K.~T., Schmidt, M.~R., et al.\ 2016, arXiv:1604.05641 

\bibitem[Loinard et al.(2012)]{2012ApJ...749L...4L} Loinard, L., Menten, K.~M., G{\"u}sten, R., Zapata, L.~A., \& Rodr{\'{\i}}guez, L.~F.\ 2012, \apjl, 749, L4 

\bibitem[Mitchell \& Robinson(1978)]{1978ApJ...220..841M} Mitchell, R.~M., \& Robinson, G.\ 1978, \apj, 220, 841 

\bibitem[Morris et al.(1999)]{1999Natur.402..502M} Morris, P.~W., Waters, L.~B.~F.~M., Barlow, M.~J., et al.\ 1999, \nat, 402, 502 

\bibitem[Prieto et al.(2014)]{Prieto2014} Prieto, J.~L., Rest, A., Bianco, F.~B., et al.\ 2014, \apjl, 787, L8 

\bibitem[Sault et al.(1995)]{1995ASPC...77..433S} Sault, R.~J., Teuben, P.~J., \& Wright, M.~C.~H.\ 1995, Astronomical Data Analysis Software and Systems IV, 77, 433 

\bibitem[Smith(2004)]{2004MNRAS.351L..15S} Smith, N.\ 2004, \mnras, 351, L15 

\bibitem[Smith(2006)]{2006ApJ...644.1151S} Smith, N.\ 2006, \apj, 644, 1151 

\bibitem[Smith et al.(2006)]{2006ApJ...645L..41S} Smith, N., Brooks, K.~J., Koribalski, B.~S., \& Bally, J.\ 2006, \apjl, 645, L41 

\bibitem[Smith et al.(2003)]{2003AJ....125.1458S} Smith, N., Gehrz, R.~D., Hinz, P.~M., et al.\ 2003, \aj, 125, 1458 

\bibitem[Smith et al.(2002)]{2002ApJ...567L..77S} Smith, N., Gehrz, R.~D., Hinz, P.~M., et al.\ 2002, \apjl, 567, L77 

\bibitem[Teodoro et al.(2016)]{2016ApJ...819..131T} Teodoro, M., Damineli, A., Heathcote, B., et al.\ 2016, \apj, 819, 131 

\bibitem[Tielens(2010)]{2010pcim.book.....T} Tielens, A.~G.~G.~M.\ 2010, The Physics and Chemistry of the Interstellar Medium, by A.~G.~G.~M.~Tielens, Cambridge, UK: Cambridge University Press, 2010,  

\bibitem[Verner et al.(2005)]{2005ApJ...629.1034V} Verner, E., Bruhweiler, F., Nielsen, K.~E., et al.\ 2005, \apj, 629, 1034 

\bibitem[Westphal \& Neugebauer(1969)]{1969ApJ...156L..45W} Westphal, J.~A., \& Neugebauer, G.\ 1969, \apjl, 156, L45 

\bibitem[White et al.(2005)]{2005ASPC..332..126W} White, S.~M., Duncan, R.~A., Chapman, J.~M., \& Koribalski, B.\ 2005, The Fate of the Most Massive Stars, 332, 129 

\end{thebibliography}
\end{document}